\documentclass[letterpaper]{article}
\usepackage{aaai2027} 

\usepackage{times}
\usepackage{helvet}
\usepackage{courier}
\usepackage[hyphens]{url}
\usepackage{graphicx}
\usepackage{amsmath, amssymb, amsfonts}
\usepackage{booktabs}
\usepackage{multirow}
\usepackage{xcolor}
\usepackage{microtype}
\usepackage{natbib}
\usepackage{algorithm}
\usepackage{algorithmic}
\usepackage{subcaption}
\usepackage{graphicx}
\usepackage{duckuments}
\usepackage{pifont}
\usepackage[table]{xcolor}
\usepackage{booktabs}
\usepackage{multirow}

\definecolor{groupbg}{HTML}{EDF1F5}   
\definecolor{oursbg}{HTML}{DFEAF8}    
\definecolor{bestbg}{HTML}{C7DCF4}    
\definecolor{inkgray}{HTML}{5A5A5A}
\usepackage{graphicx}
\newcommand{\rot}[1]{\rotatebox[origin=c]{90}{\scriptsize #1}}

\newcommand{\std}[1]{{\fontsize{6.5}{7}\selectfont\textcolor{inkgray}{$\pm$#1}}}
\newcommand{\best}[1]{\textbf{#1}}
\newcommand{\snd}[1]{\underline{#1}}

\title{BrainLinear: A Linear Model for Brain Network Analysis \\in Sparse Tangent Subspaces}

\author{
Sijing Wu\textsuperscript{\rm 1,\rm 2},
Dongyuan Li\textsuperscript{\rm 2},
Miaoting Huang\textsuperscript{\rm 1},
Weiwei Ye\textsuperscript{\rm 2},
Ying Zhang\textsuperscript{\rm 3},
Feng Xia\textsuperscript{\rm 4},
Renhe Jiang\textsuperscript{\rm 2}
}

\affiliations{
\textsuperscript{\rm 1}
Guangdong University of Foreign Studies\\
\textsuperscript{\rm 2}
The University of Tokyo\\
\textsuperscript{\rm 3}
RIKEN AIP\\
\textsuperscript{\rm 4}
Royal Melbourne Institute of Technology\\
\texttt{lidy@csis.u-tokyo.ac.jp}
}
\begin{document}

\maketitle

\begin{abstract}

Functional connectome analysis examines brain-region interactions to understand and identify disorders such as autism spectrum disorder and Alzheimer's disease. Existing methods typically use GNNs and Transformers to model the full functional connectivity matrix. However, processing tens of thousands of connections introduces redundancy and noise, increases computational cost, and limits connection-level interpretability. This raises a central question: do we really need complex interaction modeling, or is identifying a small set of disease-relevant connectivity patterns sufficient? To answer this question, we propose BrainLinear, a lightweight geometry-aware framework for mining disease-discriminative connectome patterns. BrainLinear first maps each functional connectivity matrix to a shared tangent space centered at the Fr\'echet mean of the training set, capturing subject-specific deviations while respecting matrix geometry. It then scores each ROI-pair tangent direction by its classification contribution and disease--control difference, retaining Top-$K$ directions as a compact representation. Finally, a shallow multilayer perceptron performs classification on the selected representation. Experiments on ABIDE and ADNI show that BrainLinear matches or exceeds strong GNN and Transformer baselines at a fraction of their cost: it improves AUC and ACC over the best baseline for each metric by up to $3.54$ and $1.39$ percentage points, while reducing runtime and peak GPU memory by $84.0\%$ and $68.4\%$ relative to the closest baseline in AUC. The selected directions are directionally consistent with between-group displacements and organized across major functional systems, supporting connection-level interpretation. Source code is available at: \textcolor{blue}{https://anonymous.4open.science/r/BrainLinear-3}.

\end{abstract}



\section{Introduction}



Brain disorder classification from functional connectomes models each subject as a weighted brain graph, with brain regions as nodes and edge weights encoding inter-regional functional connectivity, and uses this graph to predict disease status~\citep{liu2026spider,liu2026multiscale}. This network-level formulation is particularly well suited to complex disorders such as autism spectrum disorder and Alzheimer's disease, whose abnormalities are distributed across multiple brain regions and functional systems rather than confined to isolated regions~\cite{pagani2026autism}. By capturing these whole-brain disruptions, functional connectomes provide a quantitative foundation for objective disease assessment~\cite{lucchetti2025constructing}, mechanistic investigation~\cite{de2026genetic}, and candidate imaging biomarker discovery~\cite{ramos2026ultra}. Accordingly, functional connectome analysis has become an active research direction spanning medical image analysis, computational neuroscience, and graph learning~\cite{Yang2025BQN,ding2026syncbrain}.




Driven by growing neuroimaging datasets and advances in deep graph learning, functional connectome analysis has shifted from handcrafted connectivity features to end-to-end graph representation learning~\cite{survey2025gnnfmri, yang2026review}. Existing methods broadly fall into two categories by interaction scope: local and global interaction modeling. \textbf{For local interaction modeling}, graph neural networks propagate information across neighboring ROIs to capture disease-relevant local structure~\cite{velickovic2018gat,cui2023braingb, bessadok2022gnn,xia2026graph}. Brain-specific variants extend this paradigm by highlighting discriminative regions through ROI-aware convolution and node pooling~\cite{li2021braingnn, cui2022ibgnn}, jointly learning task-oriented connectivity structures~\cite{kan2022fbnetgen, zhang2023brainusl}, or hierarchically aggregating ROIs into disease-relevant subnetworks~\cite{xu2024contrastpool,qiu2024highorder}. \textbf{For global interaction modeling}, graph Transformers use global self-attention to capture long-range dependencies across arbitrary brain regions~\cite{ying2021graphormer, wu2021graphtrans}. Brain-specific variants further structure these global interactions by inferring latent functional modules from ROI connectivity profiles~\cite{kan2022brainnetworktransformer,bannadabhavi2023community}, jointly modeling local and long-range brain communication~\cite{yu2025causal, wei2024neuropath}, or incorporating biological priors such as small-world topology into attention~\cite{peng2025biobgt,chen2025core}.







Despite the remarkable success of existing methods, three core issues remain unresolved.
\textbf{First, ignoring the geometry of functional connectivity matrices limits predictive performance.} A functional connectivity matrix is not a collection of independently varying edge weights; its entries are coupled by global structural constraints~\cite{you2021riemannian, dan2025geodynamics}. Yet existing deep graph models typically treat it as an ordinary weighted graph and perform message passing or global attention in Euclidean space~\cite{pei2025community, wei2026large}. Although these operations capture ROI interactions, they do not explicitly preserve matrix-level structural dependencies, limiting their ability to characterize coordinated disease-related changes across multiple connections.
\textbf{Second, lacking explicit identification of discriminative connections limits interpretability.} Most methods learn over the full or nearly full connectivity space, although neuroscientific evidence suggests that discriminative information can concentrate in a limited set of connectivity patterns despite abnormalities spanning multiple functional systems~\cite{Yahata2016SmallNumber, pagani2026autism}. Explanations based on attention weights, ROI importance, or learned graph structures~\cite{yuan2025enhancing, vidya2025explainable} cannot reliably determine which connections drive classification, whether their directions align with disease--control differences, or whether they form biologically meaningful functional-system patterns.
\textbf{Third, modeling the full connectome with complex interactions incurs computational overhead.} Common brain atlases yield thousands to tens of thousands of ROI connections, yet existing methods repeatedly process this space through multi-layer message passing, graph structure learning, hierarchical aggregation, or global self-attention~\cite{dan2025holographic, chen2025core}. This incurs substantial runtime and memory costs while exacerbating redundancy and overfitting in small-sample neuroimaging settings. More fundamentally, the assumption that more extensive interaction modeling necessarily yields better brain network representations remains insufficiently tested~\cite{Yang2025BQN,han2025rethinking}.
Together, these limitations raise a central question: \textbf{how can we preserve connectome geometry and identify discriminative connections without complex full-connectome interaction modeling?}


To answer this research question, we propose BrainLinear, which replaces complex interaction modeling over the full connectome with discriminative subspace learning in a shared tangent space. \textbf{To preserve connectivity-matrix geometry}, BrainLinear treats each regularized functional connectivity matrix as a symmetric positive definite (SPD) descriptor and maps it to a tangent space anchored at the AIRM Fréchet mean estimated from the training fold, providing a common coordinate system for whole-brain connectivity variation. \textbf{To identify discriminative connections and enable their validation}, BrainLinear ranks each ROI-pair-indexed tangent direction by the product of its linear-probe coefficient magnitude and disease--control displacement magnitude, then retains the Top-$K$ directions to form a compact shared discriminative subspace. \textbf{To avoid complex full-connectome interactions}, BrainLinear trains only a shallow MLP on this compact subspace, without multi-layer message passing, graph pooling, or global self-attention. Together, these designs unify geometry-aware representation, verifiable connection-level selection, and lightweight disease classification within a single framework. Overall, our contributions are threefold:
\begin{itemize}
\item \textbf{A new paradigm for functional connectome learning.}
We replace complex interaction modeling over the full brain graph with geometry-aware discriminative subspace learning, preserving global connectome structure while retaining only compact disease-relevant information.

\item \textbf{Verifiable connection-level interpretability.} We identify discriminative connections by combining predictive relevance with group displacement, and verify that they point in the direction of the observed group differences.

\item \textbf{Evidence that lightweight modeling is sufficient.}
On ABIDE and ADNI, a shallow classifier over the selected connections outperforms
strong GNN and Transformer baselines, indicating that dense full-connectome interaction is not required for accurate
classification.
\end{itemize}

\section{Related Work}

\noindent\textbf{Deep Graph Models for Functional Connectomes.} Deep models for functional connectome analysis generally follow either message-passing GNNs or global-attention Transformers. Generic GNNs aggregate local neighborhoods with fixed or learned weights~\cite{velickovic2018gat,cui2023braingb}, while brain-specific variants add ROI-aware convolution and pooling, task-oriented network generation, and hierarchical aggregation~\cite{kan2022fbnetgen,xu2024contrastpool}. Graph Transformers instead model global dependencies through spectral, structural, or long-range attention~\cite{ying2021graphormer,wu2021graphtrans}, with brain-specific designs adding functional modules, long-range communication, and biological priors~\cite{yu2024alter,peng2025biobgt}; related work on structure-aware aggregation~\cite{yan2025hetergp,liu2025egcgn} likewise remains Euclidean. All operate in Euclidean hidden spaces over dense connectomes, without preserving matrix geometry or identifying the ROI-pair-level factors driving prediction. BrainLinear learns from a geometry-consistent tangent representation, selects and validates a compact set of discriminative coordinates, and avoids dense propagation and global attention.


\noindent\textbf{Geometry-Aware Connectome Learning.} Regularized functional connectivity matrices can be treated as SPD descriptors on a Riemannian manifold. Prior work either builds deep architectures directly on the manifold through Riemannian layers and optimization \cite{brooks2019riemannian,ho2020learning}, which retains geometry at the cost of model complexity, or maps connectomes to a shared tangent space for Euclidean learning \cite{wong2018riemannian,dadi2019benchmarking,you2021riemannian, abbas2023tangent}, keeping the full representation without task-driven coordinate selection. BrainLinear uses the tangent space for both representation and sparsification, ranking each ROI-pair-indexed coordinate by its probe relevance and disease displacement to obtain a Top-\(K\) subspace that supports compact classification and interpretation.

\begin{figure*}[t]
    \centering
    \includegraphics[width=1\textwidth]{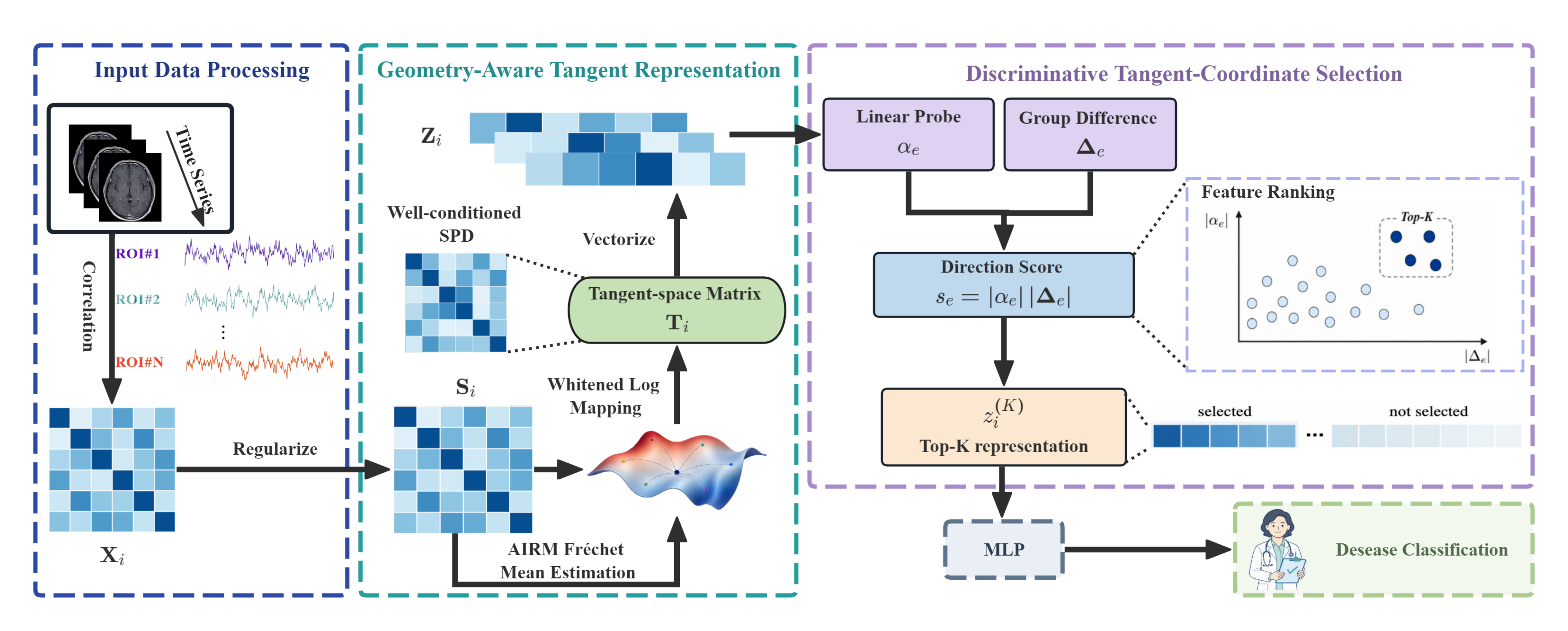}
\caption{Overview of BrainLinear. Functional connectomes are mapped to a shared AIRM tangent space, where disease-discriminative ROI-pair coordinates are selected by joint probe relevance and group displacement for lightweight classification.}
    \label{fig:overview}
\end{figure*}

\section{Problem Formulation and Notation}


Let \(\mathcal{D}=\{(\mathbf{X}_i,y_i)\}_{i=1}^{M}\) be a functional-connectome dataset, where \(\mathbf{X}_i=\mathbf{X}_i^\top\in\mathbb{R}^{N\times N}\) encodes pairwise functional connectivity among \(N\) regions of interest (ROIs) and \(y_i\in\{0,1\}\) is the diagnostic label, with \(1\) and \(0\) denoting disease and control. Our goal is to predict \(y_i\) from \(\mathbf{X}_i\). We write \(\mathcal{E}=\{(u,v):1\leq u<v\leq N\}\) for the ROI-pair index set of size \(E=N(N-1)/2\), fix a row-major bijection \(\pi:\mathcal{E}\rightarrow\{1,\ldots,E\}\) indexing ROI pair \((u,v)\) by \(\pi(u,v)\), and let \(\mathcal{I}_{\mathrm{tr}}\) denote the training-subject index set. \(\mathbb{S}^{N}\) denotes the space of real symmetric matrices, \(\mathbb{S}_{++}^{N}\subset\mathbb{S}^{N}\) the SPD cone, \(\mathbf{I}_N\) the identity matrix, and \(\|\cdot\|_F\) the Frobenius norm.

\section{Methodology}

As shown in Fig.~\ref{fig:overview}, BrainLinear comprises two stages and a lightweight classifier. It first maps each connectome to shared whitened tangent coordinates induced by the training-fold Fr\'echet mean, yielding \(\widetilde{\mathbf{z}}_i\in\mathbb{R}^{E}\). It then ranks these coordinates by discriminative relevance and retains the top \(K\) to form \(\mathbf{h}_i\in\mathbb{R}^{K}\), which is classified by a shallow MLP. Unlike dense Euclidean message passing or global attention, BrainLinear performs sparsification only after geometry-aware tangent mapping; thus, the selected features are ROI-pair-indexed tangent coordinates, rather than raw FC edges.

\subsection{Geometry-Aware Tangent Representation}
\label{subsec:tangent}

Functional connectomes are commonly represented by vectorizing the upper-triangular FC entries, which treats connections as independent Euclidean variables despite the positive-definite constraints coupling the full matrix. This representation therefore obscures coordinated changes spanning multiple connections. BrainLinear instead models each regularized connectome on the SPD manifold and linearizes it around a training-fold reference. Because the resulting tangent coordinates are derived from a matrix-level spectral mapping, they capture structured whole-connectome variation.

\subsubsection{Shared Tangent-Space Mapping}

Empirical FC matrices $\mathbf{X}_i$ are often singular or numerically ill-conditioned. We thus symmetrize each matrix and apply an adaptive diagonal shift:
\begin{equation}
\footnotesize
\mathbf{A}_i = \frac{1}{2}
\left(
\mathbf{X}_i+\mathbf{X}_i^\top
\right),\quad
\mathbf{S}_i
=
\mathbf{A}_i+\delta_i\mathbf{I}_N
\in\mathbb{S}_{++}^{N},
\label{eq:spd_descriptor}
\end{equation}
where \(\delta_i=\max\{0,\tau-\lambda_{\min}(\mathbf{A}_i)\}+\epsilon_{\mathrm{spd}}\), with \(\lambda_{\min}(\cdot)\) being the smallest eigenvalue, \(\tau>0\) being the target eigenvalue floor, and \(\epsilon_{\mathrm{spd}}>0\) denoting a small stability constant.

Since \(\mathbb{S}_{++}^{N}\) is a curved manifold, comparing descriptors requires a geometry-aware metric. We adopt the affine-invariant Riemannian metric (AIRM), whose geodesic distance is
\begin{equation}
\footnotesize
d_{\mathrm{AIRM}}(\mathbf{S}_i,\mathbf{S}_j)
=
\left\|
\log\left(
\mathbf{S}_i^{-1/2}\mathbf{S}_j\mathbf{S}_i^{-1/2}
\right)
\right\|_F.
\label{eq:airm_distance}
\end{equation}
AIRM is invariant to invertible congruence transformations, so its distance is unaffected by linear reparameterizations of the descriptors. Since tangent-space linearization is reference-dependent and less accurate for distant samples, we use the training-fold AIRM Fr\'echet mean as the common reference:
\begin{equation}
\footnotesize
\bar{\mathbf{S}}
=
\arg\min_{\mathbf{S}\in\mathbb{S}_{++}^{N}}
\frac{1}{|\mathcal{I}_{\mathrm{tr}}|}
\sum_{i\in\mathcal{I}_{\mathrm{tr}}}
d_{\mathrm{AIRM}}^2(\mathbf{S},\mathbf{S}_i).
\label{eq:airm_frechet}
\end{equation}
The minimizer is unique because the SPD manifold under AIRM is Hadamard and serves as the base point for tangent-space mapping. Then each subject is assigned the shared whitened tangent coordinates
induced by \(\bar{\mathbf{S}}\):
\begin{equation}
\footnotesize
\mathbf{T}_i
=
\log\left(
\bar{\mathbf{S}}^{-1/2}
\,\mathbf{S}_i\,
\bar{\mathbf{S}}^{-1/2}
\right)
\in\mathbb{S}^{N}.
\label{eq:tangent_mapping}
\end{equation}
The AIRM logarithmic map at \(\bar{\mathbf{S}}\) is
$\operatorname{Log}_{\bar{\mathbf{S}}}(\mathbf{S}_i)=\bar{\mathbf{S}}^{1/2}
\mathbf{T}_i
\bar{\mathbf{S}}^{1/2}.$ 
Thus, \(\mathbf{T}_i\) is the whitened coordinate representation of this tangent vector at the identity, placing all subjects in a shared fold-specific Euclidean space. The mapping satisfies
\begin{equation}
\footnotesize
\|\mathbf{T}_i\|_F
=
d_{\mathrm{AIRM}}(\bar{\mathbf{S}},\mathbf{S}_i).    
\end{equation}
Thus, it preserves each subject's AIRM distance to \(\bar{\mathbf{S}}\), although arbitrary pairwise distances between subjects are generally not preserved after linearization. The same training-fold reference is fixed when mapping validation and test subjects.

\subsubsection{ROI-Pair Tangent Feature Extraction}

Because the matrix logarithm in Eq.~\eqref{eq:tangent_mapping} is applied to the complete descriptor \(\mathbf{S}_i\), each entry \(T_{i,uv}\) is jointly determined by the full matrix rather than being a pointwise transformation of \(X_{i,uv}\). Nevertheless, \(\mathbf{T}_i\) remains symmetric and preserves the original ROI indexing, allowing each off-diagonal entry to be associated with an unordered ROI pair. Since the diagonal entries do not correspond to connections between distinct ROIs, BrainLinear retains only the strictly upper-triangular coordinates:
\begin{equation}
\mathbf{z}_i
=
\operatorname{vech}_{\mathrm{off}}(\mathbf{T}_i)
\in\mathbb{R}^{E},
\qquad
z_{i,\pi(u,v)}=T_{i,uv},
\label{eq:tangent_vectorization}
\end{equation}
where \(\operatorname{vech}_{\mathrm{off}}(\cdot)\) stacks the strictly upper-triangular entries according to the fixed row-major ordering \(\pi\).

This interpretability-oriented extraction is not an isometric vectorization of the tangent matrix because diagonal components are omitted. Each \(z_{i,\pi(u,v)}\) represents the subject's matrix-wide logarithmic displacement along the tangent coordinate associated with the ROI pair \((u,v)\); its sign denotes orientation in the shared tangent space rather than a direct increase or decrease in \(X_{i,uv}\). Before relevance estimation, each coordinate is standardized using training-fold statistics:
\begin{equation}
\footnotesize
\widetilde{z}_{i,e}
=
\frac{
z_{i,e}-\mu_e^{\mathrm{tr}}
}{
\sigma_e^{\mathrm{tr}}+\epsilon_{\mathrm{std}}
},
\qquad e=1,\ldots,E,
\label{eq:tangent_standardization}
\end{equation}
where \(\mu_e^{\mathrm{tr}}\) and \(\sigma_e^{\mathrm{tr}}\) are the empirical mean and
standard deviation of coordinate \(e\) over \(\mathcal{I}_{\mathrm{tr}}\), and
\(\epsilon_{\mathrm{std}}>0\) is a small constant. These statistics are fixed and
reused for validation and test subjects. Standardization places all coordinates on a
common scale, making probe coefficients comparable and preventing coordinate scale from
biasing relevance estimation or \(L_{2}\) regularization.

\subsection{Discriminative Tangent-Coordinate Selection}
\label{subsec:selection}

Although the shared mapping places all subjects in a common Euclidean space, the representation remains high-dimensional and difficult to interpret. We therefore rank each ROI-pair coordinate by the product of its probe coefficient and disease--control displacement, which capture conditional decision sensitivity and marginal group shift, respectively; their joint contribution better reflects diagnostically relevant separation across the two subject groups than either signal alone.

\subsubsection{Direction Relevance Estimation}

We fit an \(L_2\)-regularized logistic regression probe on the complete
standardized tangent representations. Let
\(\boldsymbol{\alpha}=[\alpha_1,\ldots,\alpha_E]^\top
\in\mathbb{R}^{E}\) and \(b\in\mathbb{R}\) denote the learned
coefficient vector and intercept, respectively. Its pre-sigmoid
decision score is
\begin{equation}
\footnotesize
f(\widetilde{\mathbf{z}}_i)
=
\boldsymbol{\alpha}^{\top}
\widetilde{\mathbf{z}}_i+b.
\label{eq:linear_probe}
\end{equation}
The probe is trained with binary cross-entropy and \(L_2\) regularization. We use its linear decision score rather than probability because the additive structure of \(f\) permits an exact coordinate-wise decomposition of the between-group score difference.

Let $ \mathcal{I}_{+} = \{i\in\mathcal{I}_{\mathrm{tr}}:y_i=1\}$ and $ \mathcal{I}_{-} = \{i\in\mathcal{I}_{\mathrm{tr}}:y_i=0\}$ denote the disease and control subjects in the training fold. Their mean standardized tangent representations are
\begin{equation}
\footnotesize
\bar{\mathbf{z}}_{\pm}
=
\frac{1}{|\mathcal{I}_{\pm}|}
\sum_{i\in\mathcal{I}_{\pm}}
\widetilde{\mathbf{z}}_i,
\qquad
\boldsymbol{\Delta}
=
\bar{\mathbf{z}}_{+}
-
\bar{\mathbf{z}}_{-},
\label{eq:group_displacement}
\end{equation}
where \(\boldsymbol{\Delta} =[\Delta_1,\ldots,\Delta_E]^\top\) and \(\Delta_e=\bar z_{+,e}-\bar z_{-,e}\) is the standardized disease--control displacement on coordinate \(e\).

Because \(f\) is affine, each group's mean probe score equals the score at its mean representation, and the intercept cancels when the two groups are compared. The resulting group-level score difference admits the exact decomposition
\begin{equation}
\footnotesize
f(\bar{\mathbf{z}}_{+})
-
f(\bar{\mathbf{z}}_{-})
=
\boldsymbol{\alpha}^{\top}
\boldsymbol{\Delta}
=
\sum_{e=1}^{E}
\alpha_e\Delta_e.
\label{eq:margin_decomposition}
\end{equation}
Accordingly, the signed contribution of coordinate \(e\) to the
between-group score difference is
\(c_e=\alpha_e\Delta_e\). We define its discriminative relevance as
the corresponding magnitude:
\begin{equation}
\footnotesize
s_e
=
|c_e|
=
|\alpha_e\Delta_e|
=
|\alpha_e|\,|\Delta_e|.
\label{eq:direction_score}
\end{equation}
The relevance score is large only when coordinate \(e\) both influences the fitted probe and exhibits a substantial disease--control displacement, capturing their joint effect beyond either \(|\alpha_e|\) or \(|\Delta_e|\) alone. Because \(s_e\) uses the magnitude of the signed contribution \(c_e\), coordinates with \(c_e>0\) or \(c_e<0\) remain eligible for selection; directional agreement between the probe coefficient and group displacement is therefore assessed empirically rather than imposed by the criterion.

\paragraph{Sparse Tangent Representation.}

Let \(\mathcal{J}=\{1,\ldots,E\}\) denote the complete tangent-coordinate index set. Given a sparsity
budget \(K\), we retain the \(K\) coordinates with the largest relevance scores. The selected index set is defined as
\begin{equation}
\footnotesize
\mathcal{J}_K=
\operatorname{TopKIdx}
\left(
\{s_e\}_{e=1}^{E},K
\right)=
\arg\max_{\substack{
\mathcal{A}\subseteq\mathcal{J}\\
|\mathcal{A}|=K
}}
\sum_{e\in\mathcal{A}}
|\alpha_e\Delta_e|.
\label{eq:topk_selection}
\end{equation}
Here, \(\operatorname{TopKIdx}(\cdot,K)\) returns the indices of the \(K\) largest scores. Equivalently, \(\mathcal{J}_K\) maximizes the total magnitude of retained coordinate-wise contributions, but not necessarily their signed sum. Writing \(\mathcal{J}_K=\{j_1,\ldots,j_K\}\) in ascending order, let \(\mathbf{P}_K\in\{0,1\}^{K\times E}\) be the selection matrix whose \(k\)-th row is \(\mathbf{e}_{j_k}^{\top}\), where \(\mathbf{e}_{j_k}\in\mathbb{R}^{E}\) denotes the \(j_k\)-th standard basis vector. The selected representation of subject \(i\) is
\begin{equation}
\mathbf{h}_i
=
\mathbf{P}_K
\widetilde{\mathbf{z}}_i
\in\mathbb{R}^{K}.
\label{eq:sparse_representation}
\end{equation}
This restricts the representation to a \(K\)-dimensional axis-aligned coordinate subspace, and since all selected coordinates lie in the same fold-specific tangent chart, every subject is described by a shared set of ROI-pair-indexed directions. The probe parameters, group displacements, and selected set \(\mathcal{J}_K\) are estimated on the training fold and fixed thereafter. Signed coordinates are retained, with \(\widetilde z_{i,e}\) positive or negative according to whether subject \(i\) lies above or below the training-fold mean along coordinate \(e\); because each coordinate stems from the matrix-wide logarithmic mapping, its sign does not imply a corresponding change in the raw FC value \(X_{i,uv}\).

\begin{table*}[!t]
\centering
\footnotesize
\setlength{\tabcolsep}{3pt}
\renewcommand{\arraystretch}{1.05}
\begin{tabular*}{\textwidth}{@{\extracolsep{\fill}} c l cccc cccc @{}}
\toprule
& \multirow{2}{*}{\textbf{Model}}
& \multicolumn{4}{c}{\textbf{ABIDE}} & \multicolumn{4}{c}{\textbf{ADNI}} \\
\cmidrule(lr){3-6} \cmidrule(lr){7-10}
& & AUC$\uparrow$ & ACC$\uparrow$ & SEN$\uparrow$ & SPE$\uparrow$
& AUC$\uparrow$ & ACC$\uparrow$ & SEN$\uparrow$ & SPE$\uparrow$ \\
\midrule
\multirow{5}{*}{\rot{GNN}}
& BrainGNN & 64.42\std{3.57} & 63.09\std{1.35} & 65.65\std{2.88}  & 60.67\std{3.68}
           & 61.81\std{1.58} & 58.72\std{3.14} & 52.88\std{9.70}  & 62.70\std{4.26} \\
& BrainGB  & 70.32\std{3.66} & 65.12\std{3.90} & 67.01\std{10.00} & 60.07\std{8.53}
           & 66.44\std{3.33} & 63.70\std{4.65} & 60.73\std{8.25}  & 64.67\std{9.07} \\
& FBNetGen & 74.50\std{3.77} & 67.09\std{3.37} & 64.71\std{9.85}  & 69.61\std{9.30}
           & 67.05\std{2.16} & 63.26\std{1.38} & 66.79\std{6.93}  & 61.31\std{9.65} \\
& GCN      & 59.59\std{3.44} & 59.30\std{3.38} & 56.67\std{4.37}  & 61.55\std{5.29}
           & 62.45\std{3.63} & 59.12\std{4.53} & 54.55\std{9.96}  & 62.00\std{8.55} \\
& GAT      & 60.43\std{3.88} & 60.10\std{4.13} & 59.26\std{5.51}  & 62.89\std{8.03}
           & 62.00\std{2.88} & 58.75\std{2.86} & 53.20\std{7.50}  & 64.79\std{7.65} \\
& ContrastPool & 57.36\std{0.87} & 57.44\std{0.69} & 57.66\std{6.85} & 57.08\std{7.79}
           & 68.17\std{3.28} & 66.21\std{3.90} & 61.51\std{7.44}
           & \cellcolor{bestbg}\best{72.43}\std{6.53} \\
\midrule
\multirow{7}{*}{\rot{Transformer}}
& ALTER    & \snd{77.99}\std{2.21} & \snd{70.10}\std{2.26}
           & \cellcolor{bestbg}\best{72.84}\std{7.40} & 67.68\std{5.81}
           & \snd{71.86}\std{3.13} & 66.92\std{3.93}
           & \cellcolor{bestbg}\best{71.55}\std{8.91} & 64.00\std{6.80} \\
& BioBGT   & 69.96\std{1.18} & 69.70\std{2.90} & 67.04\std{3.41}  & 72.02\std{4.67}
           & 63.16\std{3.74} & 62.27\std{3.23} & 63.97\std{7.88}  & 60.55\std{6.71} \\
& BrainNetTF & 77.93\std{1.41} & 69.26\std{2.26} & 65.92\std{8.60} & \snd{73.20}\std{6.06}
           & 69.73\std{2.61} & \snd{67.85}\std{2.92} & 63.64\std{6.27} & \snd{70.67}\std{8.33} \\
& Graphormer & 63.91\std{4.05} & 61.88\std{6.85} & 66.30\std{9.98} & 55.74\std{11.00}
           & 60.69\std{5.26} & 55.75\std{3.18} & 60.18\std{11.36} & 47.75\std{13.53} \\
& GraphTrans & 60.13\std{6.73} & 57.83\std{4.71} & 65.70\std{10.30} & 49.77\std{11.52}
           & 61.41\std{3.65} & 58.60\std{5.41} & 65.57\std{6.05} & 54.37\std{3.42} \\
& SAN      & 71.35\std{2.18} & 65.34\std{2.91} & 55.41\std{9.29}  & 68.39\std{7.50}
           & 66.11\std{3.41} & 61.78\std{4.22} & 53.94\std{7.56}  & 63.63\std{8.51} \\
\midrule
\rowcolor{oursbg}
{Ours} & \textbf{BrainLinear}
& \cellcolor{bestbg}\best{78.46}\std{0.38} & \cellcolor{bestbg}\best{71.49}\std{2.17}
& \snd{68.57}\std{13.51} & \cellcolor{bestbg}\best{74.23}\std{12.13}
& \cellcolor{bestbg}\best{75.40}\std{1.03} & \cellcolor{bestbg}\best{69.11}\std{1.66}
& \snd{71.22}\std{13.02} & 67.12\std{13.45} \\
\bottomrule
\end{tabular*}
\caption{Classification performance comparison.
Best results are in \textbf{bold} with shaded cells; second best are \underline{underlined}.}
\label{tab:main_results}
\end{table*}

\subsection{Lightweight Classification}
\label{subsec:classification}

Once the discriminative subspace is fixed, \(\mathbf{h}_i\) already contains
ROI-pair-indexed coordinates expressed in a shared geometric chart and selected for
group separation, so classification requires neither message passing nor global
attention over the full connectome. We apply a shallow multilayer perceptron directly
to this \(K\)-dimensional representation, where \(K\ll E\). Let \(C\) denote the
number of diagnostic classes (\(C=2\) here) and \(\mathbf{y}_i\in\{0,1\}^{C}\) the
one-hot encoding of \(y_i\); the predicted class probabilities and classification
loss are
\begin{equation}
\footnotesize
\widehat{\mathbf{y}}_i
=
\operatorname{softmax}
\left(
\operatorname{MLP}(\mathbf{h}_i)
\right),\quad
\mathcal{L}_{\mathrm{cls}}
=
-\sum_{i\in\mathcal{I}_{\mathrm{tr}}}
\mathbf{y}_i^{\top}\log\widehat{\mathbf{y}}_i,
\label{eq:prediction_loss}
\end{equation}
where \(\operatorname{MLP}(\cdot)\) is a two-layer perceptron with dropout.

\section{Experiments}


\subsection{Experimental Settings}



\paragraph{Datasets.}
We evaluate BrainLinear on ABIDE and ADNI. \textbf{ABIDE} includes 1,009 subjects (516 with autism spectrum disorder and 493 controls), represented by \(200\times200\) functional connectivity matrices constructed using the Craddock 200 atlas~\citep{craddock2012whole}. \textbf{ADNI} includes 120 subjects (47 with Alzheimer's disease and 73 controls), represented by \(100\times100\) matrices constructed using the Automated Anatomical Labeling atlas~\citep{tzourio2002automated}.

\paragraph{Baselines.}
We compare BrainLinear with 12 representative deep graph baselines, broadly
grouped according to their dominant computational mechanisms.
\textbf{GNN-based models} include the generic GCN~\citep{kipf2017gcn} and
GAT~\citep{velickovic2018gat}, as well as the brain-network-oriented
BrainGNN~\citep{li2021braingnn}, BrainGB~\citep{cui2023braingb},
FBNetGen~\citep{kan2022fbnetgen}, and
ContrastPool~\citep{xu2024contrastpool}.
\textbf{Transformer-based models} include the generic SAN~\citep{kreuzer2021san},
Graphormer~\citep{ying2021graphormer}, and the hybrid
GraphTrans~\citep{wu2021graphtrans}, together with the brain-specific
BrainNetTF~\citep{kan2022brainnetworktransformer},
ALTER~\citep{yu2024alter}, and BioBGT~\citep{peng2025biobgt}.


\paragraph{Implementation Details.}
We evaluate classification performance using \textbf{AUC}, accuracy (\textbf{ACC}), sensitivity (\textbf{SEN}), and specificity (\textbf{SPE}). We regularize each FC matrix using identity shrinkage with $\lambda=0.05$ and eigenvalue clipping with $\epsilon=10^{-6}$, and select the top $K$ tangent directions, with $K=8000$ for ABIDE and $K=2000$ for ADNI. The classifier is a two-layer MLP with hidden dimensions of $256$ and $64$ and a dropout rate of $0.5$. We train the model using AdamW with a learning rate of $3\times10^{-4}$, a weight decay of $10^{-3}$, a batch size of $64$, and at most $160$ epochs. All fold-dependent statistics and feature-selection operations are estimated exclusively from the training fold. Baseline hyperparameters follow their official implementations whenever available. We repeat each experiment across five independent runs and report the mean $\pm$ standard deviation.

\subsection{Performance Evaluation}


Table~\ref{tab:main_results} compares BrainLinear with representative GNN- and Transformer-based models on ABIDE and ADNI. BrainLinear ranks first in five of the eight dataset--metric combinations and achieves the best AUC and ACC on both datasets. 
The results further reveal three findings. \ding{182}~\textbf{The advantage holds across disorders, atlases, and sample regimes.} Baseline rankings vary substantially between cohorts. ContrastPool, for example, has the lowest AUC on ABIDE but the third-highest baseline AUC on ADNI. In contrast, BrainLinear leads AUC and ACC on both datasets despite their different diseases, cohort sizes, and parcellations. Its largest margin appears on ADNI, the smaller cohort with a lower-dimensional tangent space, suggesting that a compact discriminative subspace is valuable when subjects are limited relative to connectome dimension. \ding{183}~\textbf{Dense interaction modeling is not required for accurate connectome classification.} BrainLinear outperforms both message-passing GNNs and attention models without using either mechanism. Interaction capacity also fails to explain baseline performance: generic graph Transformers trail the brain-specific FBNetGen on both datasets, while variation within each model family exceeds the gap between families. These results suggest that domain-specific structure, captured through geometry-aware tangent mapping and discriminative coordinate selection, matters more than broader interaction modeling. \ding{184}~\textbf{The gain reflects overall discrimination, not a class-specific advantage.} ALTER achieves the best SEN on both datasets, and ContrastPool obtains the best SPE on ADNI, yet neither leads AUC or ACC. BrainLinear trails the best ADNI SEN by only 0.33 points and achieves the highest SPE on ABIDE. However, SEN and SPE vary by more than 12 points across runs, and without significance testing, the 0.47-point AUC margin on ABIDE should be interpreted as a consistent trend instead of a decisive difference.



\begin{table*}[h!]
\centering
\footnotesize
\setlength{\tabcolsep}{3pt}
\renewcommand{\arraystretch}{1.15}
\begin{tabular*}{\textwidth}{@{\extracolsep{\fill}} l cccc cccc @{}}
\toprule
\multirow{2}{*}{\textbf{Variant}}
& \multicolumn{4}{c}{\textbf{ABIDE}} & \multicolumn{4}{c}{\textbf{ADNI}} \\
\cmidrule(lr){2-5} \cmidrule(lr){6-9}
& AUC$\uparrow$ & ACC$\uparrow$ & SEN$\uparrow$ & SPE$\uparrow$
& AUC$\uparrow$ & ACC$\uparrow$ & SEN$\uparrow$ & SPE$\uparrow$ \\
\midrule
Raw FC + MLP
& 65.41\std{3.15} & 60.80\std{2.13} & 58.66\std{5.63} & 62.00\std{7.60}
& 67.47\std{13.88} & 58.33\std{12.84} & 63.33\std{35.17} & 56.38\std{16.21} \\
Full Tangent + MLP
& 77.30\std{1.31} & 67.62\std{2.49}
& \cellcolor{bestbg}\best{82.45}\std{3.41} & 53.65\std{7.43}
& 70.61\std{3.78} & \snd{64.75}\std{3.16} & 62.42\std{6.27} & \snd{67.03}\std{11.05} \\
Random-$K$ Tangent + MLP
& 72.06\std{1.83} & 63.66\std{1.24} & 74.08\std{7.99} & 53.85\std{9.81}
& 65.29\std{4.64} & 60.50\std{4.99} & 60.94\std{10.75} & 60.01\std{13.87} \\
Coefficient-only Top-$K$
& 76.48\std{1.13} & 67.92\std{3.63} & 73.27\std{10.29} & \snd{62.88}\std{16.00}
& 70.41\std{3.75} & 64.36\std{3.96}
& \cellcolor{bestbg}\best{72.37}\std{8.66} & 56.80\std{13.53} \\
Displacement-only Top-$K$
& \snd{77.68}\std{0.83} & \snd{68.61}\std{1.74} & \snd{79.80}\std{2.54} & 58.08\std{5.29}
& \snd{70.68}\std{4.41} & \snd{64.75}\std{2.70} & 69.71\std{15.58} & 60.05\std{14.69} \\
\midrule
\rowcolor{oursbg}
\textbf{BrainLinear}
& \cellcolor{bestbg}\best{78.46}\std{0.38} & \cellcolor{bestbg}\best{71.49}\std{2.17}
& 68.57\std{13.51} & \cellcolor{bestbg}\best{74.23}\std{12.13}
& \cellcolor{bestbg}\best{75.40}\std{1.03} & \cellcolor{bestbg}\best{69.11}\std{1.66}
& \snd{71.22}\std{13.02} & \cellcolor{bestbg}\best{67.12}\std{13.45} \\
\bottomrule
\end{tabular*}
\caption{Ablation results on ABIDE and ADNI (\%).
Best results are in \textbf{bold} with shaded cells; second best are \underline{underlined}.}
\label{tab:ablation}
\end{table*}

\subsection{Efficiency Evaluation}
\begin{figure}[h!]
    \centering
    \includegraphics[width=0.85\linewidth]{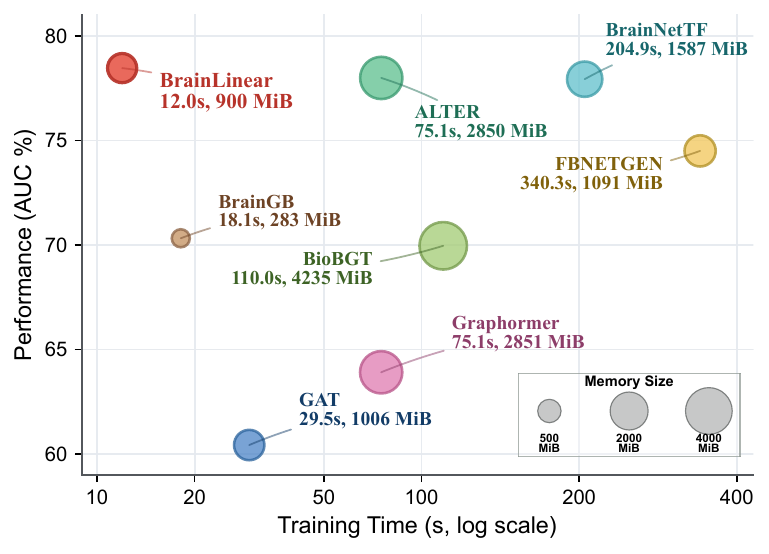}
    \caption{
    Performance-efficiency comparison on ABIDE.
    }
    \label{fig:efficiency}
\end{figure}

Figure~\ref{fig:efficiency} situates each model on ABIDE by AUC, running time, and peak GPU memory. BrainLinear sits at the upper left, attaining the highest AUC at \(12.0\)\,s and roughly \(900\)\,MiB, which cuts runtime by \(84.0\%\) and peak memory by \(68.4\%\) relative to ALTER, the closest baseline in AUC. Cost otherwise predicts accuracy poorly: BioBGT spends \(4235\)\,MiB and \(110.0\)\,s for \(69.96\) AUC and FBNetGen \(340.3\)\,s for \(74.50\), so the heaviest full-connectome models are not the most accurate. Only BrainGB is more memory-frugal ($283$\,MiB), and it trails by $8.14$ AUC points, so no baseline is simultaneously cheaper and more accurate. The reported costs follow the official baseline implementations.

\subsection{Ablation Study} 
Table~\ref{tab:ablation} isolates the contributions of tangent mapping and coordinate selection. \textbf{Raw FC + MLP} directly vectorizes the upper-triangular FC entries, whereas \textbf{Full Tangent + MLP} retains all $E$ standardized tangent coordinates. \textbf{Random-$K$ Tangent + MLP} uses a size-matched random subset, while \textbf{Coefficient-only Top-$K$} and \textbf{Displacement-only Top-$K$} rank coordinates by $|\alpha_e|$ and $|\Delta_e|$, respectively. Tangent mapping yields the largest single improvement, raising ABIDE AUC by $11.89$ points, although the corresponding ADNI gain ($3.14$) is less conclusive given the $\pm 13.88$ variance of Raw FC + MLP. Reducing the coordinate set is harmful when uninformed and neutral when guided by one signal: Random-$K$ falls $5.24$ and $5.32$ AUC points below Full Tangent + MLP, whereas ranking by $|\alpha_e|$ or $|\Delta_e|$ alone stays within $0.9$ points of it on both datasets. Only the joint criterion improves on the full representation, by $1.16$ and $4.79$ AUC points, which supports the complementary roles of the probe coefficient and the group displacement. The variants also trade specificity for sensitivity on ABIDE, where Full Tangent + MLP reaches $82.45$ SEN at $53.65$ SPE and BrainLinear is the only configuration above $70$ SPE while attaining the highest ACC on both datasets.

\subsection{Hyperparameter Sensitivity Analysis}

\begin{figure}[h!]
    \centering

    \begin{subfigure}[t]{0.48\columnwidth}
        \centering
        \includegraphics[width=\linewidth]
        {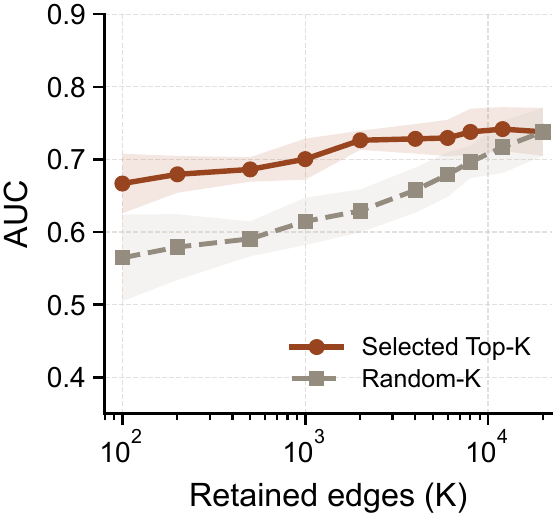}
        \caption{ABIDE}
    \end{subfigure}
    \hfill
    \begin{subfigure}[t]{0.48\columnwidth}
        \centering
        \includegraphics[width=\linewidth]
        {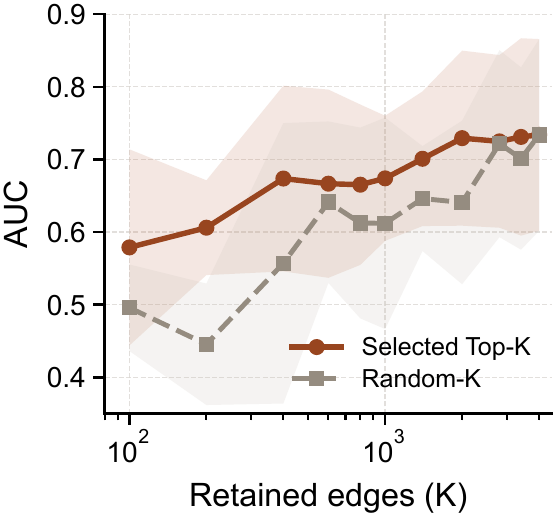}
        \caption{ADNI}
    \end{subfigure}

    \caption{Sensitivity analysis under different sparsity budgets.}
    \label{fig:k_sensitivity}
\end{figure}

We vary the sparsity budget \(K\), which fixes the dimensionality of the selected
tangent subspace. As shown in Fig.~\ref{fig:k_sensitivity}, AUC increases with
\(K\) and then saturates on both ABIDE and ADNI, so performance is not sensitive
to the exact budget once it is large enough. Across every sparsity level,
Top-\(K\) selection exceeds size-matched random-\(K\) subsets, with the gap widest
at small \(K\) and narrowing as both approach the full tangent space. Accuracy
therefore depends on which coordinates are retained rather than on how many, and
the saturation indicates that coordinates beyond the selected budget contribute
little additional discriminative signal.

\subsection{Tangent Subspace Concentration Analysis}

\begin{figure}[h!]
    \centering

    \begin{subfigure}[t]{0.48\columnwidth}
        \centering
        \includegraphics[width=\linewidth]
        {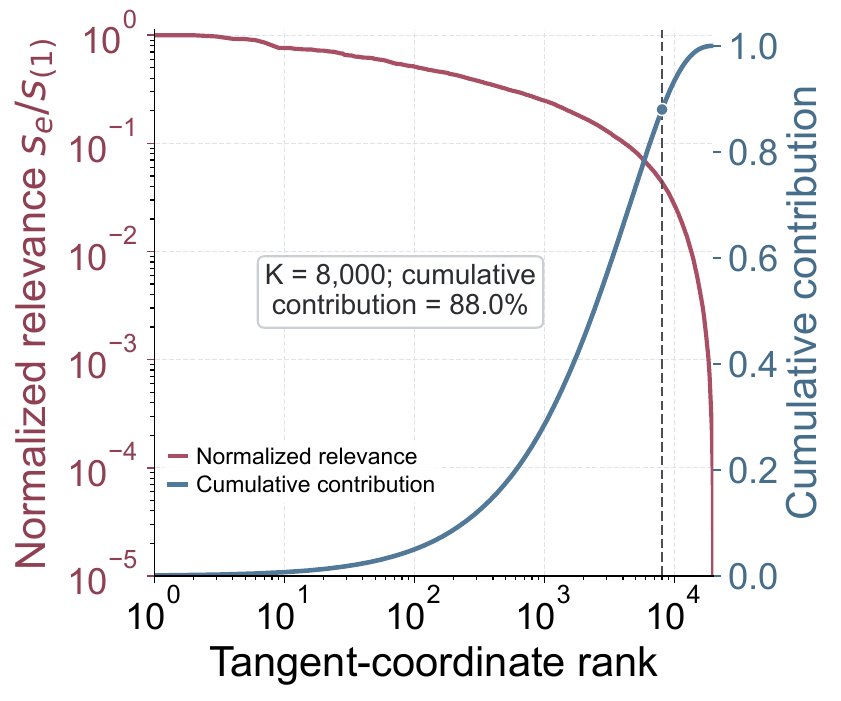}
        \caption{ABIDE (\(K=8{,}000\)).}
        \label{fig:relevance_abide}
    \end{subfigure}
    \hfill
    \begin{subfigure}[t]{0.48\columnwidth}
        \centering
        \includegraphics[width=\linewidth]
        {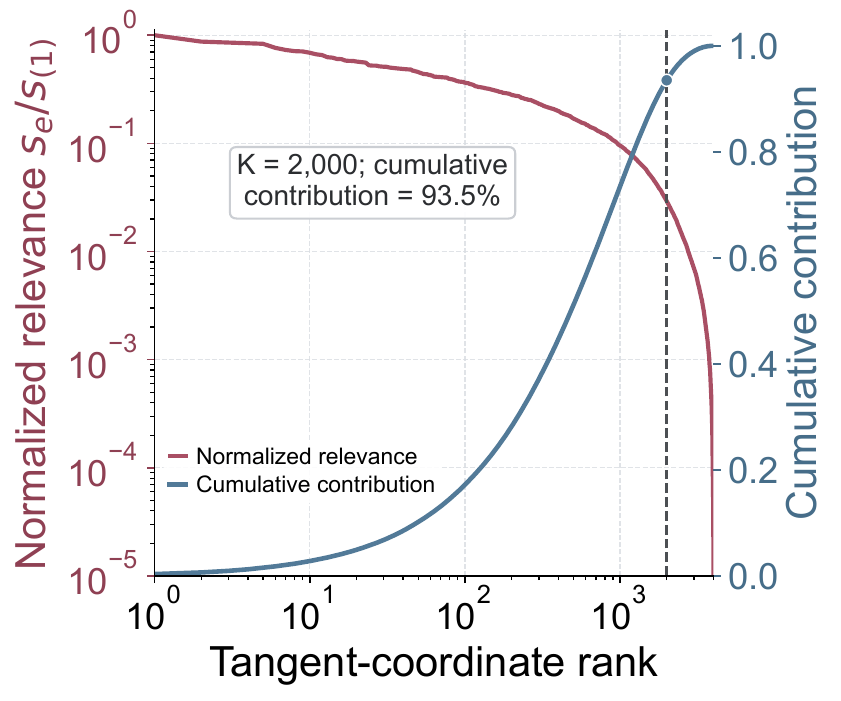}
        \caption{ADNI (\(K=2{,}000\)).}
        \label{fig:relevance_adni}
    \end{subfigure}

    \caption{
   Discriminative relevance across tangent coordinates.
    }
    \label{fig:relevance_concentration}
\end{figure}

The saturation above suggests that discriminative relevance is concentrated in a subset of tangent coordinates. We rank coordinates as $s_{(1)}\geq s_{(2)}\geq\cdots\geq s_{(E)}$ and compute their cumulative relevance. As shown in Fig.~\ref{fig:relevance_concentration}, the top $8{,}000$ of $19{,}900$ coordinates retain $88.0\%$ of the total relevance on ABIDE, while the top $2{,}000$ of $4{,}950$ retain $93.5\%$ on ADNI. The cumulative curves then plateau, showing that a compact tangent subspace captures most disease-relevant variation across both atlas resolutions while discarding largely redundant tangent coordinates that contribute little additional discriminative information.

\subsection{Interpretability Analysis}

\begin{figure}[h!]
    \centering

    \begin{subfigure}[t]{0.40\columnwidth}
        \centering
        \includegraphics[width=\linewidth]
        {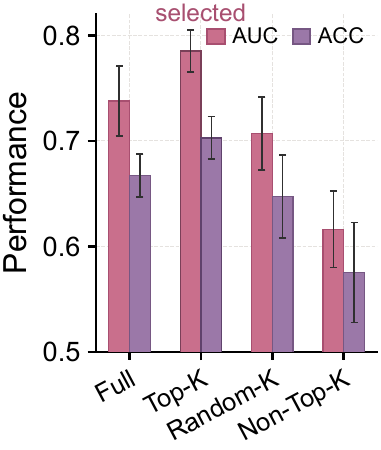}
        \caption{Performance.}
        \label{fig:selected_directions}
    \end{subfigure}
    \hfill
    \begin{subfigure}[t]{0.56\columnwidth}
        \centering
        \includegraphics[width=\linewidth]
        {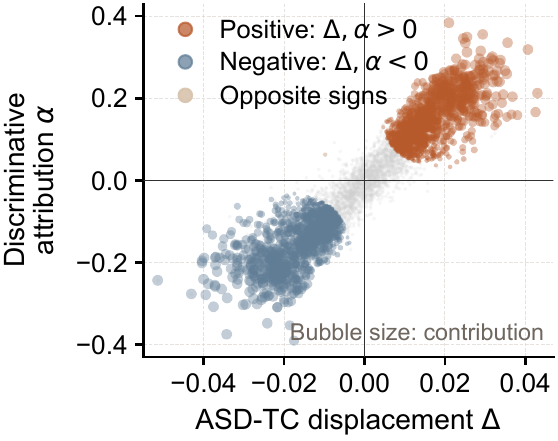}
        \caption{Directional consistency.}
        \label{fig:directional_consistency}
    \end{subfigure}

    \caption{
    Validation of selected tangent directions on ABIDE.
    }
    \label{fig:selected_direction_validation}
\end{figure}

\paragraph{Selected Tangent Direction Analysis.}
We first assess whether the Top-$K$ directions form a specific discriminative subspace instead of an arbitrary compression of the tangent representation. We use ABIDE since its larger cohort supports a more stable evaluation of sparse selection. In Fig.~\ref{fig:selected_directions}, Top-$K$ yields the highest AUC and ACC across all variants, while Non-Top-$K$ performs worst. This contrast indicates that the relevance score concentrates diagnostically useful information in the retained directions. We further compare the ASD--TC displacement, $\Delta_e=\operatorname{mean}_{\mathrm{ASD}}(\widetilde{z}_e)-\operatorname{mean}_{\mathrm{TC}}(\widetilde{z}_e)$, with the probe coefficient $\alpha_e$. Because $s_e=|\alpha_e\Delta_e|$ does not impose sign agreement, the concentration of high-contribution directions in the two sign-consistent quadrants of Fig.~\ref{fig:directional_consistency} provides empirical evidence that the fitted attributions align with the observed group-level displacements.

\begin{figure}[h!]
    \centering

    \begin{subfigure}[t]{0.46\columnwidth}
        \centering
        \includegraphics[width=\linewidth]
        {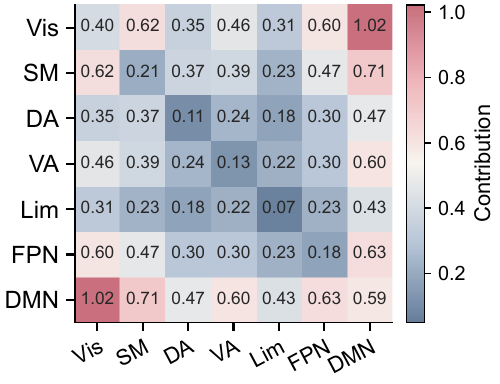}
        \caption{Functional-system contributions.}
        \label{fig:bio_system}
    \end{subfigure}
    \hfill
    \begin{subfigure}[t]{0.50\columnwidth}
        \centering
        \includegraphics[width=\linewidth]
        {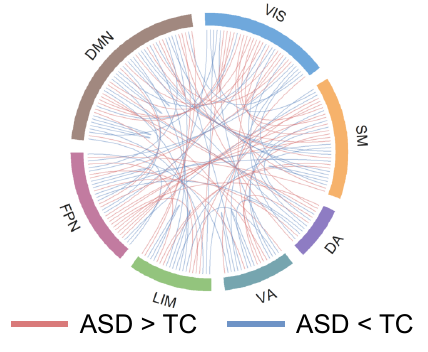}
        \caption{ROI-level connections grouped by functional systems.}
        \label{fig:bio_roi}
    \end{subfigure}

\caption{Biological interpretation of tangent directions.}
    \label{fig:bio_interpretation}
\end{figure}

\paragraph{Functional Organization Analysis.}
We examine how the selected directions are organized across functional systems and individual ROI pairs. Fig.~\ref{fig:bio_system} shows a strongly nonuniform contribution pattern, dominated by default-mode-centered interactions and connections involving the visual, somatomotor, and frontoparietal systems. Fig.~\ref{fig:bio_roi} further resolves these system-level patterns into specific ROI-to-ROI connections, including directions associated with both increased and decreased connectivity in ASD. This multiscale organization shows that BrainLinear identifies coordinated connectivity patterns spanning multiple functional systems.

\begin{figure}[h!]
    \centering

    \begin{subfigure}[t]{0.48\columnwidth}
        \centering
        \includegraphics[width=\linewidth]
        {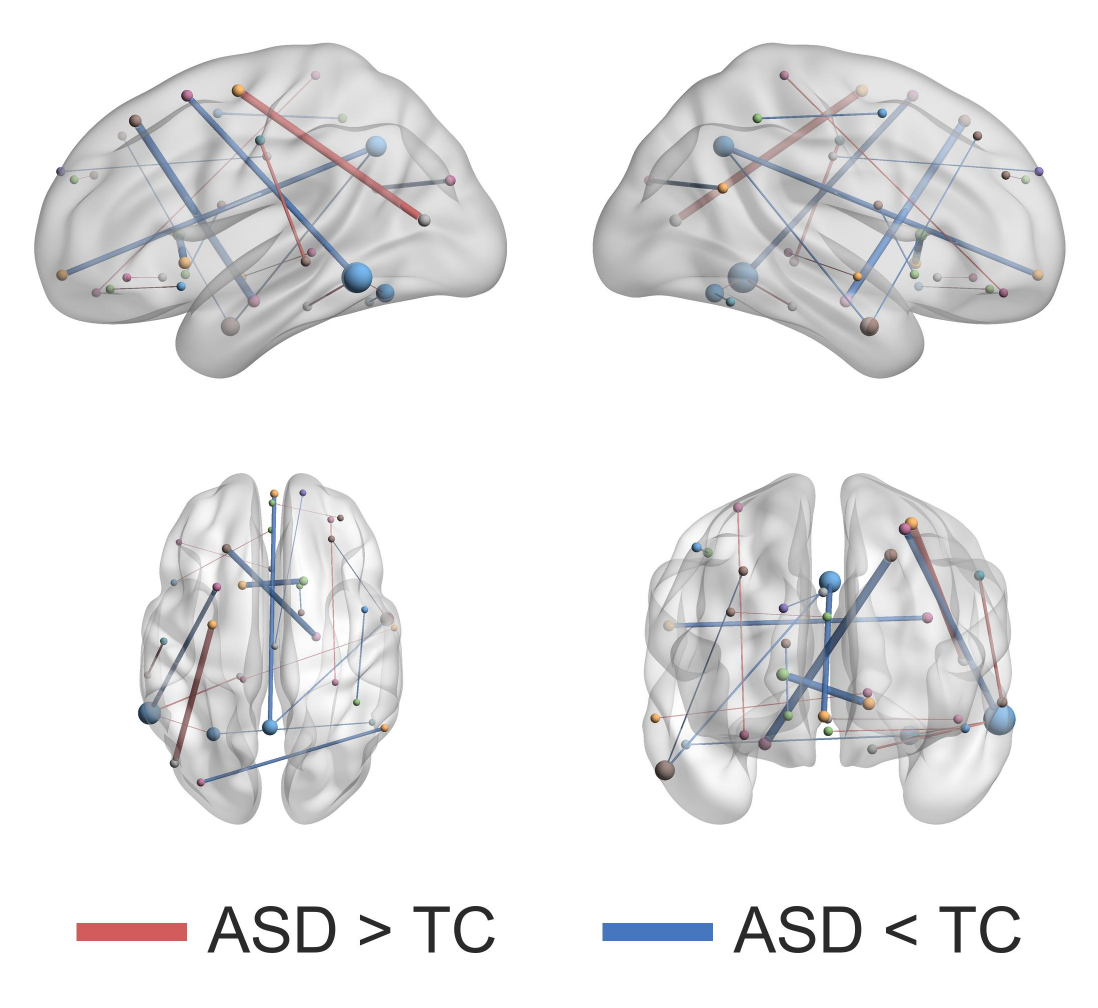}
        \caption{ABIDE.}
        \label{fig:brainnet_abide}
    \end{subfigure}
    \hfill
    \begin{subfigure}[t]{0.48\columnwidth}
        \centering
        \includegraphics[width=\linewidth]
        {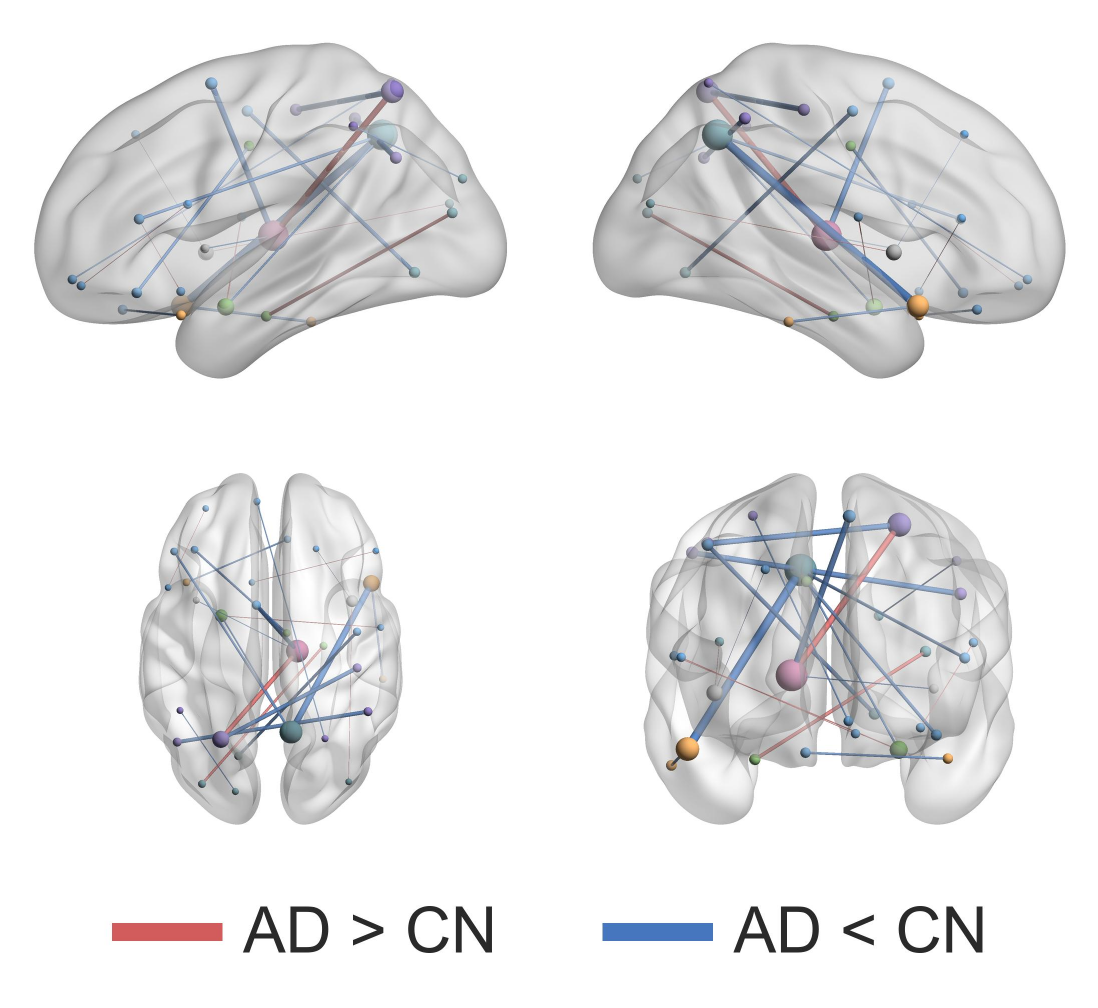}
        \caption{ADNI.}
        \label{fig:brainnet_adni}
    \end{subfigure}

\caption{Brain-space distribution of tangent directions.}
    \label{fig:brainnet_visualization}
\end{figure}

\paragraph{Brain-Space Distribution Analysis.}
We project the selected directions into anatomical brain space. In Fig.~\ref{fig:brainnet_visualization}, disease-associated connections are distributed across multiple regions and include positive and negative group displacements, supporting a network-level interpretation. The presence of distributed patterns in both cohorts provides qualitative evidence that connection-level interpretation is not limited to a single disorder. However, because these projections are descriptive, they localize model-selected patterns without establishing them as validated neurobiological biomarkers.

\section{Conclusion}


In this paper, we introduced BrainLinear, a lightweight and interpretable framework for functional connectome classification. BrainLinear maps FC matrices to a shared SPD tangent space, selects a compact set of disease-discriminative directions, and classifies them with a shallow MLP, avoiding full-connectome message passing and global attention. On ABIDE and ADNI, it outperforms strong GNN and Transformer baselines while reducing runtime and memory usage. The selected directions are non-random, aligned with between-group displacements, and organized into coherent functional-system and ROI-level patterns, indicating that discriminative information concentrates in a compact tangent subspace. Future work will extend BrainLinear to dynamic connectomes and larger prospective multi-site cohorts.

\bibliography{references}

\clearpage

\appendix
\section{Geometric Interpretation of Discriminative Tangent Selection}

The proposed representation and selection procedure can be interpreted from a
geometric perspective. Specifically, AIRM-based tangent mapping provides a
Euclidean coordinate system around the population reference point on the SPD
manifold, while discriminative selection identifies a compact subspace
formed by the most informative tangent directions.

\subsection{Isometry and Centering of the Tangent Representation}

Let $\bar{S}$ denote the Fr\'echet mean computed from the training subjects
under the affine-invariant Riemannian metric (AIRM). For each subject-specific
SPD matrix $S_i$, the whitened tangent representation is defined as

\begin{equation}
T_i=
\log
\left(
\bar{S}^{-\frac{1}{2}}
S_i
\bar{S}^{-\frac{1}{2}}
\right).
\end{equation}

\paragraph{Proposition 1 (Isometry and Centering).}
The tangent representation preserves the geodesic displacement from the
Fr\'echet mean:

\begin{equation}
\|T_i\|_F
=
d_{\mathrm{AIRM}}(\bar{S},S_i),
\end{equation}

where $d_{\mathrm{AIRM}}(\cdot,\cdot)$ denotes the affine-invariant
Riemannian distance. Moreover, the tangent coordinates are centered around the
Fr\'echet mean:

\begin{equation}
\sum_{i\in D_{tr}}T_i=0 .
\end{equation}

\paragraph{Proof.}
The AIRM distance between two SPD matrices is defined as

\begin{equation}
d_{\mathrm{AIRM}}(S_1,S_2)
=
\left\|
\log
(S_1^{-\frac12}S_2S_1^{-\frac12})
\right\|_F .
\end{equation}

By setting $S_1=\bar{S}$ and $S_2=S_i$, we obtain

\begin{equation}
d_{\mathrm{AIRM}}(\bar{S},S_i)
=
\left\|
\log
(\bar{S}^{-\frac12}S_i\bar{S}^{-\frac12})
\right\|_F
=
\|T_i\|_F .
\end{equation}

The Fr\'echet mean satisfies the first-order optimality condition

\begin{equation}
\sum_i
\mathrm{Log}_{\bar{S}}(S_i)=0 .
\end{equation}

Since

\begin{equation}
\mathrm{Log}_{\bar{S}}(S_i)
=
\bar{S}^{\frac12}
T_i
\bar{S}^{\frac12},
\end{equation}

left- and right-multiplying the above equation by
$\bar{S}^{-\frac12}$ gives

\begin{equation}
\sum_iT_i=0 .
\end{equation}

\paragraph{}
Therefore, the tangent representation provides a centered Euclidean
coordinate system while preserving the original Riemannian geometry of SPD
connectomes.

\subsection{Orthogonal Decomposition of Tangent Directions}

The tangent space of the SPD manifold is the space of symmetric matrices:

\begin{equation}
T_{\bar{S}}\mathcal{S}_{++}^{N}=Sym(N).
\end{equation}

Thus, each tangent representation can be decomposed into orthogonal basis
directions corresponding to connectivity coordinates.

\paragraph{Proposition 2 (Connectivity Directions as Tangent Basis).}
For a tangent matrix $T_i\in Sym(N)$, there exists an orthogonal basis
$\{E_e\}$ such that

\begin{equation}
T_i=\sum_e\beta_{i,e}E_e,
\end{equation}

and

\begin{equation}
\|T_i\|_F^2
=
\sum_e\beta_{i,e}^{2}.
\end{equation}

For an off-diagonal connectivity pair $(u,v)$, the corresponding basis element
is defined as

\begin{equation}
E_{uv}
=
\frac{1}{\sqrt{2}}
(e_ue_v^T+e_ve_u^T),
\quad u<v .
\end{equation}

\paragraph{Proof.}
The symmetric matrix space is equipped with the Frobenius inner product

\begin{equation}
\langle A,B\rangle_F
=
\mathrm{Tr}(A^TB).
\end{equation}

The above basis satisfies

\begin{equation}
\langle E_a,E_b\rangle_F=\delta_{ab}.
\end{equation}

Therefore, any tangent matrix admits an orthogonal expansion:

\begin{equation}
T_i=\sum_e\beta_{i,e}E_e .
\end{equation}

The squared Frobenius norm consequently satisfies

\begin{equation}
\|T_i\|_F^2
=
\sum_e\beta_{i,e}^{2}.
\end{equation}

\paragraph{}
This decomposition provides a geometric interpretation of tangent
coordinates: each ROI pair indexes a specific coordinate in the tangent space.
Importantly, this indexing correspondence does not imply an isolated
perturbation of the original functional connectivity edge, since each tangent
coordinate is jointly determined by the full SPD matrix transformation.

\begin{table*}[h!]
\centering
\footnotesize
\setlength{\tabcolsep}{3pt}
\renewcommand{\arraystretch}{1.15}

\begin{tabular*}{\textwidth}{@{\extracolsep{\fill}}
l c cccc cccc @{}}

\toprule

\multirow{2}{*}{\textbf{Feature}}
&
\multirow{2}{*}{\textbf{Classifier}}
&
\multicolumn{4}{c}{\textbf{ABIDE}}
&
\multicolumn{4}{c}{\textbf{ADNI}}
\\

\cmidrule(lr){3-6}
\cmidrule(lr){7-10}

&
&
AUC$\uparrow$
&
ACC$\uparrow$
&
SEN$\uparrow$
&
SPE$\uparrow$
&
AUC$\uparrow$
&
ACC$\uparrow$
&
SEN$\uparrow$
&
SPE$\uparrow$
\\

\midrule


\multirow[c]{3}{*}{Raw FC}

& Logistic
&65.41
&60.80
&58.66
&62.00
&66.14
&60.83
&51.56
&66.76
\\

& SVM
&68.57
&62.34
&52.35
&71.90
&62.38
&57.50
&59.56
&55.62
\\

& MLP
&69.17\std{4.11}
&63.93\std{2.31}
&54.37\std{10.22}
&73.06\std{7.90}
&67.97\std{6.49}
&59.17\std{8.25}
&63.33\std{19.85}
&55.90\std{23.15}
\\

\midrule


\multirow[c]{3}{*}{Full Tangent}

& Logistic
&73.34
&66.20
&56.38
&\cellcolor{bestbg}\best{75.58}
&67.47
&58.33
&63.33
&56.38
\\

& SVM
&73.21
&66.60
&58.41
&74.41
&68.28
&59.17
&43.11
&70.29
\\

& MLP
&77.30\std{1.31}
&67.62\std{2.49}
&82.45\std{3.41}
&53.65\std{7.43}
&70.61\std{3.78}
&64.75\std{3.16}
&62.42\std{6.27}
&67.03\std{11.05}
\\

\midrule


\multirow[c]{3}{*}{\textbf{Selected Tangent}}

& Logistic
&76.03
&64.85
&\cellcolor{bestbg}\best{85.71}
&45.19
&71.85
&66.67
&55.56
&73.33
\\

& SVM
&75.73
&63.86
&82.65
&46.15
&71.85
&\cellcolor{bestbg}\best{70.83}
&55.56
&\cellcolor{bestbg}\best{80.00}
\\

& \cellcolor{oursbg}\textbf{MLP}
&
\cellcolor{bestbg}\best{78.46}\std{0.38}
&
\cellcolor{bestbg}\best{71.49}\std{2.17}
&
\cellcolor{oursbg}68.57\std{13.51}
&
\cellcolor{bestbg}74.23\std{12.13}
&
\cellcolor{bestbg}\best{75.40}\std{1.03}
&
\cellcolor{bestbg}69.11\std{1.66}
&
\cellcolor{bestbg}\best{71.22}\std{13.02}
&
\cellcolor{oursbg}67.12\std{13.45}
\\

\bottomrule

\end{tabular*}

\caption{
Effect of classifier choice on different feature representations.
Best results are highlighted in bold with shaded cells.
}

\label{tab:classifier_choice}

\end{table*}

\subsection{Sparse Selection as Discriminative Tangent Subspace Projection}

The discriminative score used for ranking tangent directions is defined as

\begin{equation}
s_e=|\alpha_e\Delta_e|,
\end{equation}

where $\alpha_e$ denotes the classifier coefficient and $\Delta_e$ represents
the between-group difference of the corresponding tangent coordinate.

\paragraph{Proposition 3 (Top-$K$ Selection as Tangent Subspace Projection).}
Let $\mathcal{E}_K$ denote the selected Top-$K$ tangent directions according to
$s_e$. The resulting representation corresponds to a projection onto a
$K$-dimensional discriminative tangent subspace.

\paragraph{Proof.}
According to Proposition 2, the tangent representation can be expressed as

\begin{equation}
z_i=\sum_e\beta_{i,e}E_e .
\end{equation}

The selected directions define the subspace

\begin{equation}
\mathcal{V}_K=
\mathrm{span}
\{E_e:e\in\mathcal{E}_K\}.
\end{equation}

Keeping only selected coordinates is equivalent to the orthogonal projection

\begin{equation}
P_Kz_i=\Pi_{\mathcal{V}_K}(z_i).
\end{equation}

Due to orthogonality, the representation satisfies

\begin{equation}
\|z_i\|^2
=
\|\Pi_{\mathcal{V}_K}(z_i)\|^2
+
\|\Pi_{\mathcal{V}_K^\perp}(z_i)\|^2 .
\end{equation}

Hence, Top-$K$ selection provides an axis-aligned discriminative subspace in
the tangent space.

\paragraph{Remark (Population-whitened Tangent Metric).}
The feature normalization used in our method can be interpreted as a
population-adaptive rescaling of tangent coordinates:

\begin{equation}
\tilde{z}_i
=
D^{-\frac12}(z_i-\mu),
\end{equation}

where $D$ contains the variances of tangent coordinates estimated from the
training set. This operation introduces a population-whitened metric in the
tangent space:

\begin{equation}
g_{\mathrm{pop}}(x,y)=x^TD^{-1}y .
\end{equation}

Therefore, subsequent discriminative selection is performed in a
variance-adaptive tangent coordinate system.

\begin{table*}[h!]
\centering
\footnotesize
\setlength{\tabcolsep}{4pt}
\renewcommand{\arraystretch}{1.15}

\begin{tabular*}{\textwidth}{@{\extracolsep{\fill}}
c l l c c c c @{}}

\toprule

\textbf{Rank}
&
\textbf{ROI Pair}
&
\textbf{Network Pair}
&
$\Delta_e$
&
$\alpha_e$
&
\textbf{Importance}
&
\textbf{Direction}
\\

\midrule

1
&
Precentral\_L--Occipital\_Mid\_L
&
Somatomotor--Unassigned
&
0.0405
&
0.3167
&
0.0128
&
Positive
\\

2
&
Hippocampus\_R--Frontal\_Sup\_L
&
Frontoparietal--DefaultMode
&
-0.0343
&
-0.3740
&
0.0128
&
Negative
\\

3
&
Caudate\_L--Caudate\_R
&
Somatomotor--Limbic
&
-0.0514
&
-0.2431
&
0.0125
&
Negative
\\

4
&
Frontal\_Mid\_L--Temporal\_Mid\_L
&
Frontoparietal--Visual
&
-0.0431
&
-0.2761
&
0.0119
&
Negative
\\

5
&
Cerebelum\_Crus1\_R--Lingual\_L
&
Unassigned--Limbic
&
0.0355
&
0.3331
&
0.0118
&
Positive
\\

6
&
Frontal\_Med\_Orb\_L--Precuneus\_L
&
Somatomotor--Visual
&
-0.0376
&
-0.3073
&
0.0115
&
Negative
\\

7
&
Angular\_R--Occipital\_Mid\_L
&
Somatomotor--Frontoparietal
&
-0.0395
&
-0.2783
&
0.0110
&
Negative
\\

8
&
Temporal\_Mid\_L--Postcentral\_L
&
DefaultMode--VentralAttn
&
0.0293
&
0.3533
&
0.0103
&
Positive
\\

9
&
Temporal\_Mid\_L--ParaHippocampal\_L
&
Visual--Unassigned
&
0.0337
&
0.2913
&
0.0098
&
Positive
\\

10
&
Parietal\_Inf\_R--Precentral\_R
&
Limbic--Visual
&
-0.0388
&
-0.2518
&
0.0098
&
Negative
\\

\bottomrule

\end{tabular*}

\caption{
Top-ranked discriminative tangent directions identified by BrainLinear.
The importance score is defined as $s_e=|\alpha_e\Delta_e|$.
}

\label{tab:top_edges}

\end{table*}

\section{Additional Implementation Details}

\paragraph{Data preprocessing.}

All data-dependent operations are strictly performed within each training
split to avoid information leakage. Specifically, the reference point of the
SPD manifold, feature normalization statistics, discriminative direction
scores, and Top-$K$ tangent direction selection are estimated only from the
training subjects and subsequently applied to the corresponding validation and
test subjects.

To guarantee numerical stability during Riemannian computation, each
functional connectivity matrix is converted into a symmetric positive definite
matrix before tangent-space mapping. Specifically, the raw connectivity matrix
$X_i$ is first symmetrized. We then
apply identity shrinkage:
\begin{equation}
\widetilde{X}_i = (1-\lambda)X_i + \lambda I ,
\end{equation}
where $\lambda$ controls the shrinkage strength. Finally, an eigenvalue floor
is applied:
\begin{equation}
S_i = Q_i \operatorname{diag}\big(\max(d_{ij}, \epsilon)\big) Q_i^\top ,
\end{equation}
where $\widetilde{X}_i = Q_i \operatorname{diag}(d_{ij}) Q_i^\top$ and
$\epsilon=10^{-6}$ is used for numerical stability. The resulting SPD matrix
$S_i$ is used for Riemannian tangent-space mapping.

\paragraph{Optimization and model configuration.}

After discriminative tangent directions are selected, the retained tangent
coordinates are fed into a lightweight multilayer perceptron (MLP) classifier.
The classifier consists of two fully connected layers with hidden dimensions
of \textbf{256} and \textbf{64}. Dropout with probability \textbf{0.5} is
applied during training to reduce overfitting.

The model is optimized using AdamW with a learning rate of \textbf{3e-4} and
weight decay of \textbf{1e-3}. The training process is conducted for at most
\textbf{160} epochs with early stopping. For ABIDE, the final value of $K$ is
set to \textbf{8000} according to validation performance, and sensitivity
analyses over different sparsity levels are provided in the supplementary
experiments.

\paragraph{Reproducibility.}

Each experimental setting is evaluated with five independent
random initialization seeds to account for model stochasticity.
The same data partition is used across runs.
Specifically, we use five random initialization seeds, resulting in
5 runs for each experimental configuration.

The reported performance is calculated as the mean and standard deviation
over all runs. All baseline models are evaluated under their official
implementations whenever available, using the same data splits and evaluation
protocols for fair comparison.

All experiments are conducted on a workstation equipped with an NVIDIA
GeForce RTX 4060 Ti GPU with 16 GB VRAM. The implementation is based on
PyTorch, and all experiments are performed using the same computational
environment.

\section{Additional Experimental Results}

\subsection{Effect of Classifier Choice}

To examine whether the improvement of BrainLinear mainly originates from the
proposed representation rather than the classifier architecture, we evaluate
different classifiers using three feature representations: Raw FC, Full
Tangent, and Selected Tangent.

We compare Logistic Regression, SVM, and MLP under the same experimental
setting. All experiments are repeated with five random seeds. Since Logistic
Regression and SVM are deterministic classifiers, their results are reported
as single values, while MLP results are reported as mean $\pm$ standard
deviation.

The results in Table~\ref{tab:classifier_choice} show that tangent-space
representations consistently outperform raw FC features across different
classifiers. Moreover, selecting discriminative tangent directions further
improves performance, demonstrating that the gain mainly comes from the
geometry-aware representation and sparse tangent subspace rather than the
classifier choice.

\subsection{Additional Interpretability Analysis}

To further investigate whether the selected tangent directions capture
biologically meaningful connectivity patterns, we provide additional
connection-level and system-level analyses. Each selected tangent coordinate
corresponds to an ROI-to-ROI direction in the tangent space. We analyze the
most discriminative directions according to the relevance score:

\begin{equation}
s_e = |\alpha_e \Delta_e|,
\end{equation}

where $\alpha_e$ denotes the coefficient of the linear probe and $\Delta_e$
represents the between-group difference of the corresponding tangent
coordinate.

\subsubsection{Top-ranked Discriminative Tangent Directions}

Table~\ref{tab:top_edges} lists the top-ranked discriminative tangent
directions identified by BrainLinear on the ABIDE dataset. Each direction is
mapped back to its corresponding anatomical ROI pair and functional systems.
The results show that highly ranked directions are distributed across several
functional systems, including Default Mode, Frontoparietal, Visual, and
Somatomotor networks, demonstrating that the selected tangent directions
capture distributed connectivity alterations rather than isolated local
changes.

\begin{table}[h!]
\centering
\footnotesize
\setlength{\tabcolsep}{6pt}
\renewcommand{\arraystretch}{1.15}

\begin{tabular}{lcc}

\toprule

\textbf{Pattern}
&
\textbf{Directions}
&
\textbf{Importance}
\\

\midrule

$\Delta>0,\alpha>0$
&
4089
&
7.2430
\\

$\Delta<0,\alpha<0$
&
3905
&
6.9698
\\

$\Delta>0,\alpha<0$
&
4
&
0.0024
\\

$\Delta<0,\alpha>0$
&
2
&
0.0013
\\

\bottomrule

\end{tabular}

\caption{
Sign consistency analysis of selected discriminative tangent directions.
}

\label{tab:direction_consistency}

\end{table}

\subsection{Direction Consistency Analysis}

To quantify the distribution of selected tangent directions across functional
systems, we further aggregate the relevance scores according to the functional
network pair of each ROI connection. Specifically, the contribution of a
network pair $(a,b)$ is computed as

\begin{equation}
C(a,b)=
\sum_{e\in(a,b)}s_e ,
\end{equation}

where $s_e$ denotes the discriminative score of the corresponding tangent
direction.

Table~\ref{tab:direction_consistency} summarizes the contribution statistics.
The majority of selected directions belong to two dominant quadrants:
positive displacement with positive classifier coefficient and negative
displacement with negative classifier coefficient, indicating that the
selected directions consistently contribute to the classification boundary.

\section{Generative AI Use Statement}
We acknowledge the use of large language models (LLMs) solely for manuscript
preparation and language refinement. Specifically, LLMs were employed for
grammatical correction, vocabulary refinement, and sentence-level structure
improvements.

All scientific contributions, including the proposed methodology,
experimental design, data analysis, theoretical formulation, interpretation
of results, and conclusions, were developed by the authors. LLMs were only used for language polishing and manuscript editing.
\end{document}